\documentclass[10pt,preprintnumbers,nofootinbib,prd]{revtex4}

\usepackage{graphicx}
\usepackage{amsmath}
\usepackage{amssymb}
\usepackage{mathrsfs}
\usepackage{feynmf}
\makeindex \textheight=9.5in \textwidth=6.5in \topmargin=0in
\begin{document}
\title{What is the Cosmological Significance of a Discovery of\\Wimps at Colliders or in Direct Experiments?}
\author{Jacob L. Bourjaily}
\email{jbourj@umich.edu}
\author{Gordon L. Kane}
\email{gkane@umich.edu}
\affiliation{Michigan Center for Theoretical Physics,
University of Michigan, Ann Arbor, MI 48109 USA}
\date{22$^{\mathrm{nd}}$ January 2005}

\begin{abstract}
Although a discovery of wimps either at colliders or in direct experiments would have enormous implications for our understanding of particle physics, it would imply less than one would like about our understanding of the dark matter in the universe or in the galactic halo: it surely is possible that the discovered particles account for only a little of the total dark matter. To establish the cosmological significance of a wimp discovery, their density must be determined. We show that data from neither hadron colliders nor direct detection experiments alone can be sufficient to determine the local or relic density of discovered wimps, even allowing all needed assumptions about cosmology and astrophysics. We provide examples of dark matter models where nearly identical detector or collider signals correspond to very different densities. We show, however, that it may be possible to determine the density of wimps by combining data from both experiments; we present a general method to do this in the case of supersymmetric dark matter, and describe how similar studies could be made for other wimp candidates.\\

\end{abstract}

\maketitle
%\newpage
%\tableofcontents\thispagestyle{empty}
\newpage
\section{Introduction}\vspace{-0.3cm}
\indent We are now confident that our universe contains a large amount of cold dark matter. The most popular particle candidates for dark matter are stable, weakly interacting massive particles (wimps). These particles are being searched for directly and indirectly by many experimental groups throughout the world, and may be produced at colliders in the near future. But we must ask, what is the {\it cosmological} significance of a discovery of wimps? The answer depends on what fraction of the total dark matter they compose.\\
\indent If wimps are discovered experimentally, there are many reasons to hope (and perhaps expect) that they solve the dark matter problem---{\it i.e.} that the relic density\footnote{Throughout this paper we will try to distinguish between the {\it cosmological} and {\it local} (halo) relic densities of wimps. Although both of these are `relics' of the early universe, we will exclusively use `relic density' for the `cosmological' relic density. In Section II.C.1, we describe why such care is necessary.} of the wimps is equal to the known relic density of cold dark matter. A universe in which almost all the dark matter was composed of a single kind of particle would be one arguably kind to physicists. But of course, whether or not we live in such a universe is a question that can only be asked and answered experimentally.\\
\indent Throughout our discussion, we will find that the connection between the density of discovered wimps and the data seen at colliders or in direct detection experiments is quite tenuous: it will be all too easy to find models which agree with the initial data having almost any desired relic density. To only consider models consistent with the data that solve the dark matter problem is to avoid the physical question of the nature of dark matter. To use the known relic density of dark matter to constrain the relic density of a particular wimp candidate is to make unjustified claims based on wishful thinking. Put another way, cosmological data does not imply a lower bound on the relic density of any  particular wimp; there exists only an upper bound: no wimp can compose more than the entirety of dark matter.\\
\indent The question of the composition of dark matter is usually avoided by the implicit assumption that only one wimp would exist in nature. Of course this hope would need to be verified experimentally. But even if there were only one wimp particle in nature, there would still be no reason to assume that its relic density were equal to that of cold dark matter. This is because wimps are not the only kind of particle that can contribute to the total dark matter of the universe. For example, there are strong theoretical motivations for the existence of axions; if axions exist, they may account for a significant fraction of the total cold dark matter. But it may not be possible to accurately compute the amount of axions produced (non-thermally) in the early universe \cite{Brhlik:2000dm}. It may be that the only way to determine what (possibly significant) fraction of dark matter is axionic is to determine the relic abundance of all other contributions. Therefore, the only way to solve the dark matter problem, will be to use experimental data to `measure' the actual relic abundance of discovered wimps. As we will find, this could only be (and perhaps can be) done using data from both colliders and direct experiments combined.\\
\indent Although the production of wimps at colliders would have enormous implications for our understanding of elementary particle physics, it implies less than one would like about dark matter. As it turns out, it may be exceedingly difficult to calculate their (thermally-produced) relic density with data from hadron colliders for general wimp models. This is even true for specialized models, such as supersymmetric dark matter. We give examples of models with nearly indistinguishable collider signatures that have vastly different relic densities. Furthermore, there is no reason to expect that all the dark matter was produced thermally, or that thermal production took place `canonically' in the early universe---{\it e.g.} in the absence of quintessence. We must therefore directly measure their density in the galactic halo, and this cannot be done at colliders. But even if these issues could be dealt with, the needed particle physics information is unlikely to exist. \\
\indent Fortunately, data from direct detection experiments is directly related to the actual local density of wimps in the galactic halo. However, to extract the local density from these experiments, the wimp must be identified and its couplings to matter must be calculated. And this information is only obtainable at colliders. Until the wimp is observed and studied at colliders, the local density will be invisible to direct experiments. These issues, for colliders and direct detection were first raised technically although not resolved in Ref \cite{Brhlik:2000dm}. Again, we give examples of models with nearly indistinguishable in/direct detection signals that have very different relic densities.\\
\indent To determine the local halo density of a new particle, its scattering rate with matter must be observed in direct detection experiments and its cross sections computed using data from colliders. But this requires that the wimp produced in colliders is in fact the same as that observed directly. If the particles observed in each had the same mass, it would (strongly) suggest that they were the same. We show that it may be possible to determine the wimp mass using data from direct detection experiments alone, and discuss the difficulties of this measurement at hadron colliders.\\
\indent Once the local density of a new particle is determined, we can deduce what fraction of the local dark matter it represents. This will experimentally establish the relevance of the particle to cosmology. Furthermore, analysis of this type could lead to novel and strong limits on high-scale physics including the existence of additional, high-scale gauge symmetries or quintessence dark energy.\\
\indent This paper is organized as follows. First, we study how hadron colliders and dark matter experiments can study dark matter independently. We show by explicit examples that, in contrast to what is suggested by some specialized studies, it will be extremely difficult (impossible?) to determine the relic abundance of generic dark matter wimps produced thermally in the early universe using hadron colliders alone. We emphasize that even if the thermal production calculation were possible, there would remain very serious uncertainties regarding non-thermal (and non-standard) production.\\
\indent We study how dark matter wimps in the galactic halo could be observed directly or indirectly in experiments. We explain why---and demonstrate with explicit examples---that no combination of data from these experiments alone can determine the local density of wimps unless the particle has been identified and its scattering cross sections known. To calculate the coupling of a discovered wimp to the matter in detectors, one must know many details of its particle physics; of course this requires that it be identified. This information can only be obtained at colliders. We present a general method to determine the local density of wimps: using direct detection experiments to measure the product of the scattering cross section and their local density, and colliders to identify the particle and estimate its cross sections. We discuss important astrophysical caveats of this line of reasoning, and describe possible ways for these issues to be resolved.\\
\indent In order to explicitly show how direct detection data depends on the particle physics and local density of a wimp, we briefly review how dark matter particles are detected in these experiments. This will allow us to introduce formalism common in the literature, and show how data from experiments with different detector materials and at different recoil energies can be combined to improve our understanding of dark matter in the halo. This discussion will naturally motivate a general (and possibly robust) technique to determine the mass of particles seen in these experiments---which is critical to its identification.\\
\indent We then consider (in detail) the case where the wimp seen in experiments is the lightest supersymmetric particle (LSP) predicted by minimally supersymmetric extensions of the standard model (MSSMs). We derive a robust upper bound for (what amounts to) the LSP-nucleon scattering cross section within the framework of the most general MSSM using only sparse data from colliders. This analysis applies to the most general softly-broken MSSM with a neutralino LSP; no presumptions are made about supersymmetry breaking, the masses of sparticles, or mixing phases. We show how this bound can be combined with direct detection data to determine an absolute lower bound on the local density of neutralino dark matter \cite{Bourjaily:2004aj,Bourjaily:2004fp}.

\section{The Relic and Local Densities of Wimps: Colliders and Detectors}%\vspace{-0.4cm}
\indent There is an analogy to be drawn relating our understanding of the dark matter problem to our attempts to solve it: the two principal sources of evidence for the existence of dark matter in the universe being cosmology and astrophysics; the two ways in which the problem is hoped to be solved being collider physics and in/direct detection experiments.\\
\indent Studies of the cosmic microwave background and the formation of large scale galactic structure indicate that the universe contains an enormous amount of cold (hence massive) matter particles---an amount which greatly exceeds that of visible matter in the universe. The excess is called `cold dark matter' (cdm). Given in terms of its ratio to the critical energy density of the universe multiplied by Hubble's constant $h$, the `relic density' of cold dark matter is confined to the range \mbox{$0.094<\Omega_{\mathrm{cdm}}h^2<0.129$} \cite{Bennett:2003bz}. This is the (universal) relic density of all cold dark matter, whatever it is composed of; popular components of dark matter include old stellar material, axions, and the lightest supersymmetric particle---all of these and perhaps more making up the total density.\\
\indent Independent of (and perhaps more forceful than) these cosmological studies, it has been known for more than seventy years that vast `halos' of invisible matter are necessary to explain the anomalous dynamics of galactic clusters and galaxies---including our own. For example, studies of the rotation of the Milky Way in the vicinity of the sun show that the local density of our galactic halo is roughly $\sim0.3$ GeV/cm$^3$ \cite{Caldwell:1981rj,Gates:1995dw}. This matter in the halo, (perhaps) appropriately also named dark matter, is not confined to the planes of galaxies and so must interact at most weakly (and gravitationally) with ordinary matter and itself. This  fact is strengthened by the non-existence of exotic heavy isotopes, which implies that the halo dark matter does not interact strongly or electromagnetically\footnote{If electrically or strongly charged dark matter existed in the halo, it would likely bind to nuclei causing the appearance of `exotic' heavy isotopes.} \cite{Wolfram:1979gp}. Because both cosmological and astrophysical data imply the existence of cold dark matter, the two are supposed to be the same. In short, we know the local density of dark matter and we know the cosmological density of dark matter; but the sources of these two facts are quite distinct.\\
\indent Similarly, there exist two general approaches to solve the dark matter mystery: producing dark matter particles at colliders and directly observing them in the galactic halo. Surely it would be very exciting to discover wimps in the halo or produce them at colliders. But how does the existence of wimps account for the dark matter in the universe? Certainly the mere existence of such particles implies little about cosmology. To authoritatively solve the dark matter problem, one must authoritatively demonstrate that discovered particles have the correct local or relic densities. It is not sufficient to show that there exists a model in agreement with data in which the wimp composes all of the dark matter in the universe; such analyses---likely to follow the discovery of wimps---merely avoid the mandate of determining the density of discovered particles. Like the case of cosmology and astrophysics, colliders and direct searches offer distinct, crucial clues required to solve the puzzle.

\newpage

\subsection{The (Thermal) Production of Dark Matter in the Early Universe and at Colliders}%\vspace{-0.3cm}
\indent The popularity of the supposition that dark matter is composed of wimps is the result of a complicated mixture of circumstantial evidence (together with a bit of intuition, hope, and the impressive ease of proposing new wimp models---as is evident by a glance at the literature). Experimentally, data from cosmology (including large-scale structure formation simulations) indicate that dark matter is massive (cold) and non-baryonic; data from astrophysics (including searches for exotic heavy isotopes) indicate that dark matter is at most weakly interacting. That dark matter would be well explained by `wimps' is readily encouraged by a rough estimate of what their relic density today would be from thermal production in the early universe.\\
\indent In the hot early universe, the production and annihilation of wimps would have been in equilibrium. When the universe cooled to an ambient temperature too low for their production, equilibrium would be lost and their density would decay exponentially by annihilation. This would continue until the universe expanded to such an extent that wimps would no longer effectively `find each other' to annihilate, and the (comoving) density would `freeze-out.' A back-of-the-envelope calculation for a generic wimp---using a weak-scale mass and annihilation cross section and a slightly non-relativistic average velocity---is extremely encouraging: one obtains $\Omega_{\mathrm{wimp}}h^2\sim0.1$, the cosmological result\footnote{For a slightly more explicit demonstration of this derivation see {\it e.g.} Ref. \cite{Munoz:2003gx}.}. This calculation should not be considered as encouraging as it may at first appear---and a negative result should not have been discouraging---because proper calculations for explicit models can give relic densities that vary over several orders of magnitude.\\
\indent In general, the amount of thermally-produced dark matter in the universe is calculable for any explicit model. If colliders observe wimps, then it is {\it this} way in which it is (na\"{i}vely) hoped to establish what fraction of dark matter they compose: if one knows (or guesses) enough of the particle physics describing the wimp---its mass and many of other lagrangian parameters---then one can straightforwardly compute the thermal contribution to its relic abundance.\\
\indent There are at least two important shortcomings of using hadron collider data to estimate the thermally-produced contribution to the relic abundance of wimps: one is practical (and perhaps unsurmountable); the other is one of principal. We describe each below in turn.

\subsubsection{Why the Thermal Relic Density Cannot be Determined with Hadron Colliders Alone}
\indent Unfortunately, the number of parameters needed to reliably calculate the thermal contribution to the relic abundance of wimps is typically so large that there is little hope of ever knowing the result. It should be emphasized that in general the relevant parameters are meaningful, physical quantities that are in principle measurable once the relevant new physics can be studied experimentally. But the inherent difficulty of obtaining the relevant high-precision measurements at hadron colliders makes the task appear unsurmountable: the calculation crucially relies on precise knowledge of too many lagrangian parameters. For example, in the case of a supersymmetric wimp (the lightest supersymmetric particle), virtually none of the tens of required parameters are known to be measurable in a model-independent way at hadron colliders. Therefore, although one would hope to use collider data to determine at least the thermally-produced contribution to the relic density, this may be a practical impossibility.\\
\indent To demonstrate the loss of generality typically made for the sake of progress, consider the case of supersymmetry---certainly the most actively studied model for dark matter. The most general softly-broken MSSM lagrangian contains more than one hundred currently unmeasured parameters, many of which enter the calculation of the relic density of the LSP. All of these are eventually measurable when superpartners and their interactions can be studied experimentally---but a linear collider may be necessary to achieve this. To make the problem tractable, most researchers chose to greatly reduce the number of unknown parameters by making a number of somewhat {\it ad hoc} assumptions about the model, such as how supersymmetry is broken. Virtually all of the existing work on supersymmetric cold dark matter has been done in the context of minimal supergravity, mSUGRA, which reduces the hundred unknown parameters to four and one sign.\\
\indent But even in the extremely restricted framework of mSUGRA, it may not be possible to determine the relic density in general; and there exists very little research on general mSUGRA models. Usually, the mSUGRA parameter space is further restricted to those narrow regions in which the LSP is {\it all} of the dark matter---{\it i.e.} the thermally-produced contribution to the relic density of the LSP (denoted $\chi$) is such that $\Omega_{\chi}=\Omega_{\mathrm{cdm}}$. However, even in this extremely constrained subset of all possible, extremely simplified (mSUGRA) supersymmetric models, there are regions in which the computation of the relic density of the LSP is not known to be possible. For example, mSUGRA models in the `higgs pole' region with large $\tan\beta$ suffer from enormous computational difficulties \cite{Battaglia:2004mp}.\\
\indent Perhaps (not too) surprisingly, we have found the relic density calculated within the framework of mSUGRA not to be representative of supersymmetric models even slightly more general than mSUGRA. Contrary to assumptions common in the literature, the mSUGRA results are not even indicative of models where the LSP is almost entirely Bino\footnote{The Bino is the superpartner of the standard model $U(1)$ gauge boson---a superposition of the photon and $Z$. A characteristic of mSUGRA models is the tendency for an almost purely-Bino LSP.}. For example, it is argued that in mSUGRA the masses of squarks are usually irrelevant to the calculation of the relic density, but that a small change in the mass of the LSP can have a dramatic effect \cite{Battaglia:2004mp}. This suggests that the relic density of $\chi$ crucially depends on $m_{\chi}$ and not so much on the other parameters of the model.\\
\indent However, when the assumptions of mSUGRA are relaxed even slightly, the correlation between $m_{\chi}$ and the relic density virtually disappears, as many more parameters become important.\begin{figure}[t]\includegraphics[scale=0.9493]{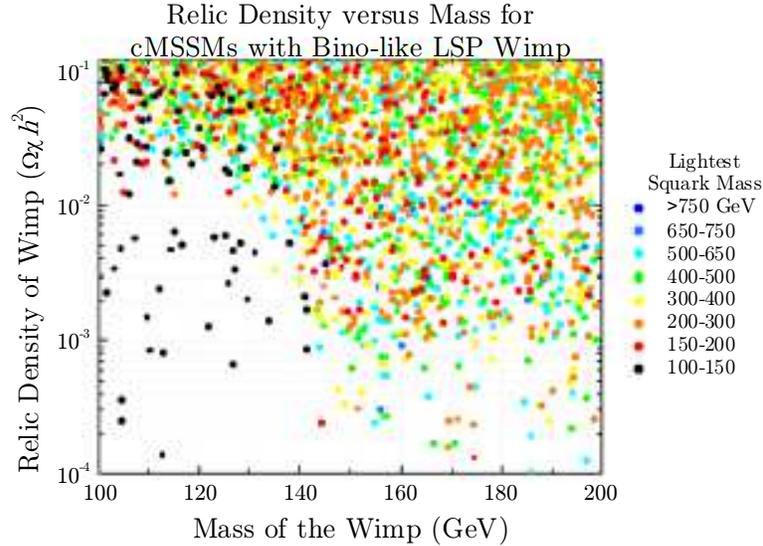}\caption{The relic densities of more than 2500 constrained MSSMs with a Bino-like LSP as a function of $m_{\chi}$. The mass of the lightest squark is indicated by color. Notice that there is almost no correlation whatsoever between mass and relic density. The lack of correlation is clear even when one knows the mass of the lightest squark.}\label{density_versus_mass}\end{figure} \mbox{Figure \ref{density_versus_mass}} shows the relic density versus LSP mass for a few thousand randomly generated, experimentally constrained\footnote{By `constrained,' we mean that all of the models are allowed under all current experimental constraints on supersymmetry and so are physically realizable. We must emphasize that this may not be the same as frequently-used definitions of `constrained' susy models in the literature. Many authors, for example, consider `constrained' to refer to {\it exclusively} mSUGRA models. In addition, some authors further add the restriction that $\Omega_{\chi}=\Omega_{\mathrm{cdm}}$.} minimally supersymmetric standard models which have a Bino-like LSP\footnote{By `Bino-like,' we mean that the Bino component of the LSP is $\geq97\%$.} (no presumptions were made about the supersymmetry breaking scenario). The mass of the lightest squark in each model is indicated by color. Notice that there is almost no correlation between the LSP mass and relic density. Furthermore, the lack of correlation is unchanged even if the mass of the lightest squark were also known. Apparently many more parameters must be known to determine $\Omega_{\chi}h^2$.\\
\indent Throughout this work, we will make use of a collection of several thousand randomly-generated, constrained MSSMs. Each was generated using the DarkSUSY code \cite{Gondolo:2004sc}, and are consistent with all known experimental constraints\footnote{These include limits on $b\to s\gamma$ and $(g-2)_{\mu}$ (the early summer of 2004 bounds), together with the requirement that the thermal contribution to the relic density be no more than that of all the dark matter. We have not tuned the parameters to stringently satisfy all the constraints on ($m_h$, \&etc), since these should not affect the general results.}. Each model is given in terms of seven (high-scale) parameters, which are:
\begin{itemize}
\item $\mu$, the higgsino mass parameter;
\item $m_{1/2}$, the (assumed universal, high-scale) gaugino mass parameter;
\item $m_A$, the mass of the CP-odd higgs particle;
\item $\tan\beta$, the ratio of the vacuum expectation values of the two higgs doublets;
\item $m_0$, the (assumed universal, high-scale) squark mass parameter;
\item $A_b$, the bottom-sector trilinear coupling;
\item $A_t$, the top-sector trilinear coupling.
\end{itemize}
Although these models are extremely simplified MSSMs (having reduced the more than one hundred parameters to seven), they do not rely on any particular supersymmetry breaking mechanism and are less constrained than mSUGRA models.\\
\indent Presumably, calculations done within mSUGRA are not very likely to precisely reflect reality (unless the real world were in fact mSUGRA, which is unlikely, and also would be very hard to demonstrate indeed). Therefore, it will be necessary to work within a more general framework than that of mSUGRA. But we have found that MSSMs even slightly more general than mSUGRA indicate that it may not be possible to determine the relic density of the LSP for a general MSSM using data from hadron colliders.\\
\begin{table*}[t]\caption{\label{examples_for_detectors}Examples of minimally supersymmetric standard models, consistent with experimental constraints, in which the (neutralino) LSP is the wimp dark matter particle candidate.}
\begin{tabular}{lccccccc}
\hline\hline& $\mu$ \footnotesize{(GeV)} & $m_0$ \footnotesize{(GeV)} & $m_{1/2}$ \footnotesize{(GeV)} & $m_A$ \footnotesize{(GeV)} & $\tan\beta$ & $A_t/m_0$ & $A_b/m_0$\\
\hline {\bf Model 1.a}\hspace{0.5cm} & $~~308.2$ & $413.3$ & $327.3$ & $380.2$ & $11.04$ & $-0.9846$ & $-0.9842$\\
\hline {\bf Model 1.b} & $-441.5$ & $405.5$ & $326.3$ & $337.6$ & $15.07$ & $~~0.8234$ & $~~0.0224$\\\hline\hline
\end{tabular}\end{table*}
\indent Consider for example the two (mSUGRA-like) MSSMs listed in \mbox{Table \ref{examples_for_detectors}}. In \mbox{Table \ref{examples_for_colliders2}} we list the particle spectra for each model together with the (thermal) relic density of each. The two models differ in relic density by a factor of more than 50: the LSP of Model $1.$a would account for all of the dark matter while that of Model $1.$b would account for $\lesssim2\%$\footnote{Contrary to the intuition from mSUGRA, notice that the model with the smaller gaugino fraction has the larger relic density.}. Despite the vast difference in their relic densities, the models have nearly identical spectra. These represent a proof by example that it will be very difficult if not impossible to compute the relic density using optimistic data at hadron colliders even for an MSSM slightly more general than mSUGRA. And in practice, this would need to be done for MSSMs much more general than those we studied.
\begin{table*}[t]\caption{\label{examples_for_colliders2}Particle spectra for example susy Models $1.$a \& $1.$b. Notice that although these models differ in relic density by a factor of 50, their signals at the LHC are virtually identical. All masses are in GeV.}
\begin{tabular}{lccccccccc}
\hline\hline& ~\hspace{0.3cm}$\Omega_{\chi}h^2$ \hspace{0.3cm}~& $m_{\chi}$ & $m_{\mathrm{NLSP}}$ & $m_{\tilde{t}}$ & $(m_{\tilde{u}}-m_{\tilde{t}})$ & $m_{h}$ & $m_H$ & $m_{\tilde{g}}$ & gaugino fraction \\
\hline {\bf Model 1.a}\hspace{0.2cm}& $0.09905$ & $158$ & $263$ & $351$ & $58.7$ & $112$ & $381$ & $1138$ & $0.935$\\
\hline {\bf Model 1.b} & $0.00172$ & $163$ & $311$ & $360$ & $42.2$ & $110$ & $338$ & $1134$ & $0.986$\\\hline\hline
\end{tabular}\end{table*}

\indent Although the models have strikingly similar spectra at hadron colliders, they give vastly different dark matter detection signals. In \mbox{Table \ref{examples_for_colliders3}} we list representative data that would be seen in direct and indirect dark matter detection experiments. Notice that the difference in density is quite apparent in their relative signal strengths. As described later in this work, the direct detection signals are roughly a measure of the product of the local density and the scattering cross section. It is natural that models with similar particle spectra would have similar scattering cross sections; such models should be differentiable easily in (and perhaps only in) dark matter detection experiments.
\begin{table*}[t]\caption{\label{examples_for_colliders3}Direct and indirect detection data for example susy Models $1.$a \& $1.$b. Notice that although these models have virtually identical particle spectra, their direct detection signals differ dramatically.}
\begin{tabular}{lcccc}
\hline\hline&  ~\hspace{0.3cm}$\rho_{\chi}$\footnote{We assume the relic and local densities to be proportional.} \hspace{0.3cm}~& Ge signal\footnote{As observed in a 20 keV bin, in units of cpd/kg-keV. An isothermal halo model is assumed.} & NaI signal\footnote{As observed in a 5 keV bin, in units of cpd/kg-keV. An isothermal halo model is assumed.} & Solar $\nu$ flux\footnote{In units of muons/yr-km$^2$.}  \\
\hline {\bf Model 1.a}\hspace{0.2cm} & $0.2971$ & $1.18\times10^{-5}$ & $4.07\times10^{-5}$ & $7.67\times10^{-1}$\\
\hline {\bf Model 1.b}               & $0.0052$ & $3.14\times10^{-8}$ & $1.07\times10^{-7}$ & $9.13\times10^{-4}$ \\\hline\hline
\end{tabular}\end{table*}
\indent Based on our results above, we expect that it is not possible to calculate the thermally-produced contribution to the relic density of a generic wimp using data from hadron colliders---even within the framework of a specific wimp model ({\it e.g.} supersymmetry). Recall, however, that there were {\it two} general shortcomings of this approach---what we have just described is the practical difficulty of obtaining enough of the parameters to make the computation. {\it Even if it were possible}, however, there remain serious shortcomings of principal.

\subsubsection{The Non-Thermal (and Non-Standard) Production of Wimps in the Early Universe}
\indent Thermal production in the early universe is the simplest possible mechanism to explain today's abundance of dark matter; it is by no means the only mechanism. There is no reason to suspect that all of the dark matter was produced thermally, and any non-thermal contribution to dark matter would yield thermal production estimates to be not only misleading but quite inadequate. The cosmological data which indicates today's abundance of dark matter is indifferent to the mechanism by which dark matter particles were produced, and there is no known way to determine what fraction of the dark matter was produced thermally.\\
\indent Perhaps the most commonly considered form of non-thermal production is the decay of heavy particles or condensates into wimps. There are many known examples of non-thermal production mechanisms for supersymmetric (LSP) dark matter. For example, inspired by anomaly-mediated supersymmetry breaking (AMSB) scenarios, heavy gravitinos produced during reheating could decay into LSPs, creating a cosmologically significant LSP relic density \cite{Gherghetta:1999sw}. Alternatively, moduli fields (flat directions, {\it e.g.} from string theory) which acquire mass from supersymmetry breaking could decay into large amounts of LSP dark matter before big-bang nucleosynthesis \cite{Moroi:1999zb}. Other mechanisms include the late-time decay of Q-balls produced in {\it e.g.} Affleck-Dine baryogenesis (see {\it e.g.} Ref. \cite{Fujii:2002kr}), or the decay of cosmic strings formed via spontaneous breaking of  high-scale $U(1)$ gauge symmetries \cite{Jeannerot:1999yn}. These mechanisms could be very important because in many viable supersymmetric models ({\it e.g.} AMSB), thermal production of the LSP would be insufficient to account for the dark matter in the universe alone.\\
\indent In fact, several non-thermal production mechanisms, including presently unknown ones, could have occurred during our cosmological history. And virtually all of these mechanisms critically rely on high-scale physics well beyond the reach of foreseeable colliders. Therefore, these mechanisms cannot be excluded using collider data and will always exist as possible alternatives to thermal production. Hence, there is no reason to suspect that the thermally-produced contribution to the relic density of wimps is the same as the {\it actual} relic density of dark matter measured cosmologically. These credible objections to thermal freeze-out calculations cannot be answered with colliders only sensitive to the effective, low-scale physics. \\
\indent But the questions of non-thermal wimp production are by no means the only objections to thermal calculations. As it turns out, even thermal production can be very sensitive to high-scale physics well beyond the reach of colliders. For example, scalar-tensor quintessence dark energy could dramatically alter the relic density resulting from thermal production. As shown in Ref. \cite{Catena:2004ba}, the modified expansion rate of the early universe in quintessence models can enhance the relic abundance of wimp dark matter by as much as three orders of magnitude. Therefore, high-scale physics could have an enormous effect even on the amount of dark matter that is produced thermally in the early universe.\\
\indent Therefore, if we hope to account for all the dark matter in the universe we cannot rely on calculations of thermal freeze-out. There is no known way to eliminate the possibility of non-thermal production or high-scale effects which cannot be accounted for using colliders in the foreseeable future. All of the arguments above indicate that it is not possible to definitively determine the relic density of dark matter particles using data from hadron colliders alone. At best, with the most optimistic data and less than general models, one could hope to compute the `canonical' thermally-produced component of the cosmological relic density. But because of the effects of quintessence, even this calculation cannot be considered conclusive. \\
\indent The only way to meet these challenges is to measure the actual {\it local} density of wimps and compare this with the known local density of dark matter in the halo. This of course cannot be done at colliders. If the local density of wimps could be determined, then dark matter could be accounted for in spite of all the ambiguities associated with non-thermal production and high-scale physics. Nevertheless, though colliders cannot do the job alone, they will be essential to the determination of the actual relic density, as we describe below.\\
\indent It has not escaped the authors' attention that {\it if} it is determined that all of the dark matter in the halo is composed of the wimps produced in colliders, then non-thermal production mechanisms could be enormously constrained by making specific (although perhaps unjustified) correlations between the local and relic densities of wimps. We will discuss this further in Section C.1.

\subsection{Directly Interacting with Wimps in the Local Halo}
\indent As we have seen, the dark matter problem can be partially addressed with colliders only after overcoming enormous obstacles of practicality and ignoring critical objections of principal. At best, colliders alone can be used to {\it calculate} the {\it cosmological} amount of dark matter wimps that {\it may have been} produced {\it thermally} in the early universe. This should be strongly contrasted with dark matter direct detection experiments, which hope to {\it measure} the {\it actual} {\it local} density of dark matter wimps {\it in our galactic halo today}.\\
\indent First proposed by Goodman and Witten \cite{Goodman:1985dc}, dark matter wimps in the halo could be observed directly through their occasional scattering with ordinary matter in earth-based detectors. Extremely sensitive calorimeters record the small amount of recoil energy deposited by wimp scattering. Because wimps are expected to have masses on the order of a hundred GeV and move relatively slowly in the halo---on the order of a few hundred km/s---they would deposit recoil energies up to $\sim 200$ keV. Direct detection experiments measure this recoil, and record the wimp-nucleus scattering rate as a function of recoil energy and time.\\
\indent The scattering rate observed in these experiments is a measure of the product of the local wimp density and the wimp-nucleon scattering cross section. If one knew the scattering cross section, then these signals directly determine the local density of particles. And because we know from astrophysical studies that the local density of dark matter is roughly $\rho_{\mathrm{cdm}}\sim0.3$ GeV/cm$^3$, the local density of wimps directly determines the fraction of the local dark matter they compose. It must be noted, however, that there are some important caveats involved in this line of reasoning, and these will be discussed below in Section C.1\\
\indent But because the scattering cross section is required to compute the local density from direct detection data, colliders will be absolutely critical. Only using data from colliders can the cross section be calculated to determine the local density. We will illustrate that this could be done through example and explicit calculations.

\subsubsection{Why the Local Density of Wimps Cannot be Determined with In/Direct Detection Alone}
\indent Let us imagine that a weakly interacting massive particle $\chi$ has been unambiguously observed in dark matter detection experiments. Because the interaction rate of wimps with a detector is a measure of the local density multiplied by the scattering cross section, low-density wimps with a high cross section are indistinguishable from high-density wimps with a low cross section. But it is precisely these two extremes that are produced by `thermal freeze-out' in the early universe: if the cross section is large then more wimps would have annihilated in the early universe and resulting density would be small; alternatively, if the cross section is small then freeze-out would have occurred very early and the density today would be high. This crude argument suggests that even a very small component of dark matter may have a detectable signal: it would likely have a relatively large cross section. This well-known result has been referred to as the `no-lose theorem' in recent work and conferences. Indeed, experimentalists may not lose out on discovering even a tiny fraction of the dark matter \cite{Duda:2002hf,Bourjaily:2004aj,Bourjaily:2004fp}. But `no-lose'$\neq$win.\\\begin{figure}[t]\includegraphics[scale=0.8507]{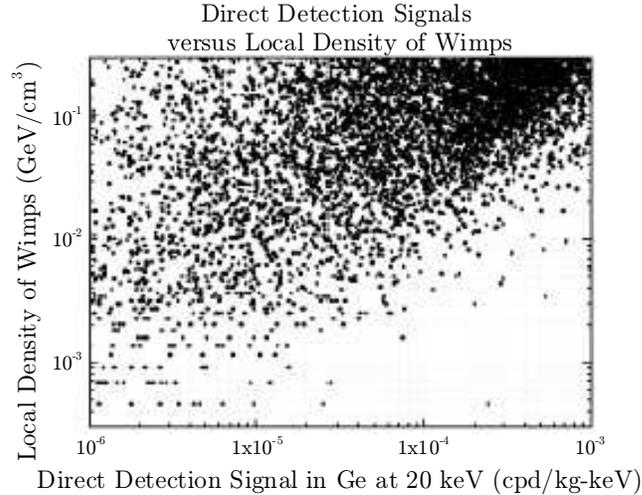}\caption{The local halo densities of several thousand constrained MSSMs as a function of direct detection signal strength (in germanium). To generate this plot, the galactic and cosmological fractions of wimp dark matter were assumed to be same---see Section C.1 for a discussion of the implicit caveats of this line of reasoning. Experiments currently in planning or under construction may be able to observe signals of the order $10^{-4}$ cpd/kg$\cdot$keV. Notice that any particular signal strength is consistent with a large range of local densities.}\label{direct_detection}\end{figure}
\indent To see how little the density of wimps is correlated to our ability to detect them, we explored randomly-generated (physically allowable) supersymmetric dark matter models. In \mbox{Figure \ref{direct_detection}}, we plot the local density\footnote{This data was generated assuming that the local density is proportional to the cosmological density.} against a representative direct detection signal for some six thousand constrained minimally supersymmetric standard models (without assuming any supersymmetry breaking scenario). The lack of correlation is immediately obvious. Notice that for any particular signal strength the local density fluctuates over at least two orders of magnitude. Hence, in accordance with the no-lose theorem, the discovery of wimps in the galactic halo tells us very little about how much of the local dark matter they compose. \mbox{Figure \ref{direct_detection}} indicates that a wimp discovery in the near future could easily represent a negligible fraction of the local dark matter, even $1\%$ or less.\\
\indent Because direct detection signals are universally proportional to the local wimp density, it is clear that no combination of direct detection data could resolve the density---there would be no data independent of this parameter. What is not as obvious, however, is that indirect detection data may not resolve the density either. There are two principal methods of indirectly detecting wimps in the halo: the study of high energy neutrinos from the annihilation of wimps in the sun or earth and the study of cosmic rays. The flux of high energy neutrinos from wimp annihilation in the sun or earth offers little more insight into the local density of wimps because---considering the age of the solar system---the wimp annihilation rate is controlled by the wimp-capture rate \cite{Jungman:1995df}; this is essentially a measure of the elastic scattering rate, and so cannot be used to separate the local wimp density from the scattering cross section any more than direct detection experiments.\\
\indent The study of the galactic gamma-ray spectrum or the anti-matter excess in cosmic rays---although not directly related to scattering rates---cannot be used to determine the density of a wimp for different reasons. Specifically, the total flux of particles from wimp annihilation in the halo is proportional to a local, unknown `boost factor'\footnote{The `boost factor' is related to galactic confinement times and other astrophysical issues. It collectively represents many (typically local) astrophysical uncertainties.} which can enhance the flux of antiparticles and gamma rays by one or two orders of magnitude (for a recent review see \mbox{Ref. \cite{Bertone:2004pz}}). Therefore, these indirect searches are sensitive to the shape of the wimp density distribution, but insensitive to the actual local density of wimps.\\
\indent From our discussion above, it should not be surprising if the local density of wimps cannot be determined in general from any collection of dark matter detection experiments. Indeed, we have found many examples where models with quite different relic densities show nearly identical signals for every type of direct and indirect detection experiments.
\begin{table*}[t]\caption{\label{examples_for_detectors6}Examples of minimally supersymmetric standard models, consistent with all current constraints, in which the LSP is the wimp dark matter particle candidate.}
\begin{tabular}{lccccccc}
\hline\hline& $\mu$ \footnotesize{(GeV)} & $m_0$ \footnotesize{(GeV)} & $m_{1/2}$ \footnotesize{(GeV)} & $m_A$ \footnotesize{(GeV)} & $\tan\beta$ & $A_t/m_0$ & $A_b/m_0$\\
\hline {\bf Model 2.a}\hspace{0.5cm} & $~~463.4$ & $593.8$ & $371.1$ & $428.2$ & $18.39$ & $-0.1844$ & $0.4366$\\
\hline {\bf Model 2.b}               & $-412.1$  & $435.5$ & $372.7$ & $337.4$ & $23.33$ & $-0.0119$ & $0.0707$\\\hline\hline
\end{tabular}\end{table*}
Consider the two (mSUGRA-like) supersymmetric models listed in \mbox{Table \ref{examples_for_detectors6}}. We calculated the in/direct detection signals and relic densities for each model using the DarkSUSY code \cite{Gondolo:2004sc}; these are listed in \mbox{Table \ref{examples_for_detectors2}}. Notice that although the two models give rise to virtually indistinguishable in/direct detection data, they differ substantially in relic density.\\
\indent Also listed in \mbox{Table \ref{examples_for_detectors2}} is the mass of the scalar up-squark for each model. Although it could be measured only at colliders, we list $m_{\tilde{u}}$ among these experiments for a purpose that will be better understood later in this work. As we will see, the masses of the squarks are critically related to the scattering cross section. Heuristically, scattering is typically dominated by squark exchange and is roughly proportional to $1/(m_{\tilde{q}}^2-m_{\chi}^2)$. Therefore, two models with the same detection signals will have densities such that \begin{equation}\frac{(\rho_{\chi})_1}{(\rho_{\chi})_2}\sim\frac{(m_{\tilde{q}}^2)_1-(m_{\chi}^2)_1}{(m_{\tilde{q}}^2)_2-(m_{\chi}^2)_2}.\end{equation} This back-of-the-envelope analysis works remarkably well for the two models, as can be easily verified numerically.\\
\begin{table*}[t]\caption{\label{examples_for_detectors2}Direct and indirect detection data for example susy Models $2.$a \& $2.$b. Notice that the two models have nearly identical direct and indirect detection signals although their relic densities differ significantly.}
\begin{tabular}{lccccccc}
\hline\hline&~\hspace{0.3cm}$\rho_{\chi}$\footnote{We assume the relic and local densities to be proportional.} \hspace{0.3cm}~& $m_{\chi}$ (GeV)& Ge signal\footnote{As observed in a 20 keV bin, in units of cpd/kg-keV. An isothermal halo model is assumed.} & NaI signal\footnote{As observed in a 5 keV bin, in units of cpd/kg-keV. An isothermal halo model is assumed.} & Solar $\nu$ flux\footnote{In units of muons/yr-km$^2$.} & Galactic $\bar{p}$ flux\footnote{Measured at $3$ GeV, in units of $\bar{p}/$cm$^2$-sr-s-GeV.}  & $m_{\tilde{u}}$ (GeV) \\
\hline {\bf Model 2.a}\hspace{0.2cm} & $0.2856$ & $184$ & $3.50\times10^{-6}$ & $1.18\times10^{-5}$ & $0.121$ & $4.58\times10^{-10}$ & $591.3$\\
\hline {\bf Model 2.b}               & $0.1044$ & $185$ & $4.09\times10^{-6}$ & $1.38\times10^{-5}$ & $0.106$ & $9.92\times10^{-10}$ & $432.2$\\\hline\hline
\end{tabular}\end{table*}
\indent Ours and similar studies suggest that no combination of dark matter in/direct detection data can be used to determine the local density of wimps alone. Therefore, it is critical that wimps also be observed in colliders. By producing wimps at colliders and studying their particle physics, the wimp-nucleon scattering cross sections may be calculated. Combining the scattering cross sections with direct detection data would immediately give the local density.

\newpage
\subsection{Solving the (Local) Dark Matter Problem}
\indent Our studies suggest that the dark matter problem can only be solved in general if wimps are observed and studied at both colliders and direct detection experiments. In Section A, we showed that colliders could at best estimate the (quintessence-free) thermally-produced component of the cosmological relic density of wimps. But these analyses require enormous lack of generality through oversimplification (such as assuming mSUGRA in supersymmetric models) and rely on high-scale physics which could not be observed at colliders (such as the non-existence of quintessence or additional high-scale gauge symmetries). In contrast, direct detection experiments are sensitive to the {\it actual} local density of dark matter wimps. But these experiments cannot measure the local wimp density without first knowing the relevant scattering cross sections---and this can only be determined using collider data.\\
\indent To determine the cosmological significance of wimps observed in both direct detection experiments and at colliders, we must measure their density in the galactic halo. This can be done by combining scattering cross sections determined using collider data with data direct detection experiments. But first it must be established that the particles observed in the two experiments {\it are in fact the same}\footnote{Furthermore, it must be established that direct detection experiments observe only a single type of wimp particle: maybe multiple wimp components exist in the halo. We are unaware of any concrete way of establishing this fact, other than perhaps observing consistency in the mass calculations described in this work.}. There is only one known way to {\it suggest} that the two signals correspond to the same wimp: by showing that each has the same mass. Fortunately, it may be possible to determine the mass of particles observed in each type of experiment independently.\\
\indent Perhaps surprisingly, it appears much easier to determine the mass of wimps using direct detection experiments than with colliders. This is because there exist model-independent ways of determining the wimp mass with direct detection data alone; these methods are described in Section III.B. At least one of these methods, such as that presented herein, could always be used to determine the mass (in principal). In strong contrast, there exists no general procedure to determine the mass of wimps produced at hadron colliders. This is because, being weakly interacting and stable, wimps indicate their presence as missing energy only: mass differences are easily measurable, but masses themselves require intensive statistical analysis. Such statistical techniques have been developed to determine the mass of the LSP for some types of mSUGRA models (see {\it e.g.} Ref. \cite{Bachacou:1999zb}). However, there is no known way to measure the LSP mass in the general MSSM. It may not even be possible to determine the mass of an arbitrary wimp produced at colliders in a model-independent way. Even in the (albeit extremely favorable) situations of some supersymmetric models where kinematics alone allows one to determine the LSP mass, the methods require enormous amounts of data. Therefore, it could be many years before there is a determination of a wimp mass from hadron colliders\footnote{Of course, soon after signals of new physics are reported at hadron colliders, such mass determinations will be reported; but they will likely depend on many restrictive, unconfirmed assumptions (such as mSUGRA). But solving the dark matter problem is too important to risk these types of misguided errors.}. But such a determination is crucial to the identification of the wimp seen in direct detection experiments.\\
\indent After the particle has been identified and its mass is known, then collider data can be used to estimate (or bound) the wimp-nucleon scattering cross section; this allows one to determine the {\it actual} local density of wimps. In contrast to the `thermal' relic abundance calculations described above, it is found that only minimal data from colliders is required to {\it absolutely bound} the scattering cross section for {\it general} (supersymmetric) models. Therefore, even sparse collider data could immediately translate into strong bounds on the actual local wimp density.\\
\indent We will illustrate how these estimates are made in the framework of a supersymmetric (neutralino) wimp for the most general softly-broken MSSM. We will see that not only can the scattering cross section be bounded strongly using even minimal data, but also that this bound rapidly approaches the true value as more collider data is obtained. These conclusions are true without relying on any assumptions about supersymmetry breaking, relative sparticle masses, or other restrictions on the MSSM.\\
\indent Combining estimates or bounds of the wimp-nucleon scattering cross sections from colliders with direct detection data immediately allows one to estimate or bound the actual local density of wimps. Up to some caveats described below, this `measured' local density of wimps would allow one to determine what fraction of cold dark matter they represent, thereby addressing their cosmological significance. To the extent that it is determined that $\rho_{\chi}$, measured with colliders and direct detection experiments, is equal to $\rho_{\mathrm{cdm}}$, measured by the orbital velocity of nearby stars in the Milky Way, the dark matter problem will be solved.

\subsubsection{Caveats and Ambiguities About the Local Density of Wimps}%\vspace{-0.3cm}
\indent Because direct detection experiments are sensitive to local, small-scale structure in the galactic halo, our existing knowledge of the ambient halo density may not be sufficient to solve the dark matter problem in general. Our knowledge of the local halo density---that is, the fact that $\rho_{\mathrm{cdm}}\sim0.3$ GeV/cm$^3$---is based on relatively large-scale surveys of star velocities in the Milky Way; these measurements are not very sensitive to small-scale structure in the halo such as local density perturbations.\\
\indent Most of the small-scale structure considered in the literature involves local, high-density regions of dark matter within the halo. Although these structures would make it easier to discover dark matter, they make it nearly impossible to assess what fraction of the halo is composed of a discovered wimp. (But of course, we could be in a low-density region.)\\
\indent There are several types of small-scale halo structure which may effect direct detection experiments. For example, some authors have proposed that the halo may be clumpy or contain caustic structures such as rings or shells (see {\it e.g.} Refs. \cite{Moore:2001vq,Sikivie:1997ng}). Alternatively, the earth may be within a stream of dark matter. This situation has been suggested by studies of the Sgr A galactic tidal stream; it has been estimated that this stream may increase the local halo density by $0.3-23\%$ relative to the ambient density \cite{Freese:2003tt}. These small-scale perturbations in the dark matter density could have significant effects on direct detection rates---and nearly fatal effects on our ability to resolve the composition of dark matter.\\
\indent Fortunately, there may exist ways to eventually exclude or compensate for most of these perturbations. For example, dark matter streams or caustics would produce anisotropy in the local wimp velocity profile; these may be identified using directional detection experiments like DRIFT \cite{Munoz:2003gx}. A clumpy dark matter halo may possibly be identified by studying the time-dependence of a wimp signal\footnote{It may not be possible to observe fluctuations in the scattering rate caused by a clumpy halo, considering the time scales which may be involved. These types of questions will not have definite answers until more work has been done to understand the physics of realistic clumpy halo models.}. These ambiguities will need to be addressed before the dark matter problem has been put to rest. Fortunately, because these obstacles could possibly be surmounted in the future, they are much better than those associated with non-thermal (and non-standard) production in the early universe. A determination of the actual local density of dark matter particles is as close to a cosmological result as we can get.\\
\indent Unfortunately, there does not exist any well-understood or accepted physical argument which allows one to go from the the local halo density to the cosmological relic density or {\it vice versa}. The relationship between the two involves many details of galaxy formation and structure that remain to be understood. Recall that the two principle sources of evidence for the existence of dark matter, cosmology and astrophysics, are both independent and distinct: it is a hopeful assumption that the dark matter required by the cosmic microwave background and large scale structure formation is precisely (and entirely) the same material that is required by the dynamics of galaxies. Therefore, even if all the ambiguities about the local halo structure were resolved and all the local dark matter was identified, it not obvious that all of the dark matter in the universe has been accounted for. However, these ambiguities would not prevent our complete understanding of the composition of the Milky Way's dark matter halo. Of course, we do expect that cosmology and astrophysics are measuring the same dark matter, but the issue is so important that this should not be automatically assumed.
%\vspace{-0.3cm}
\subsubsection{Constraining the High-Scale Physics of Non-Thermal Wimp Production}%\vspace{-0.3cm}
\indent Recall that there exists many non-thermal mechanisms to produce wimps in the early universe or enhance the number of wimps thermally produced. {\it If} and when the local density of wimps is established by the procedure described above, we would be tempted to infer the relic density by associating the relic density of dark matter, $\Omega_{\mathrm{cdm}}h^2\sim0.1$, with the local density of dark matter, $\rho_{\mathrm{cdm}}\sim0.3$ GeV/cm$^3$. {\it If} this assumption were made, then up to the caveats described in the previous section, one could infer the actual {\it relic} density of wimps. As described above, this is {\it not} a trivial assumption.\\
\indent And {\it if} there existed sufficient data to approximate the amount of wimps produced thermally in the early universe, this could be compared to their inferred relic density. Such analyses could enormously constrain the existence of many forms of high-scale physics. For example, an agreement between the two results would imply that all of the dark matter was indeed produced thermally and without enhancement from quintessence. Alternatively, any discrepancy between the two would imply the need for new physics at the high-scale. This type of analysis would be a unique window to high-scale physics.\\
\indent Therefore, after wimps are discovered and their local density determined, thermal freeze-out calculations may improve our understanding of the very same physics which today prevents freeze-out calculations from helping us understand dark matter.

\section{When Wimps are First Discovered: Making the Most of Direct Detection Data}%\vspace{-0.4cm}
\indent As we have shown, direct detection data is critical to solving the dark matter problem. When data first becomes available, it will be necessary to determine the wimp mass and separate the various independent contributions to the scattering rate. To elucidate this procedure, we will briefly review the standard formalism for the direct detection signal. This will allow us to demonstrate methods to optimize data, and to introduce an explicit technique to determine the wimp mass. This discussion will also introduce the formalism used later in this work to bound the LSP-nucleon scattering cross section.
%\vspace{-0.2cm}
\newpage
\subsection{The Scattering Rate of Halo Wimps with a Solid-State Detector}%\vspace{-0.4cm}
\indent Let us imagine that the wimp $\chi$ has been observed in direct detection experiments. It will be helpful for us to review the explicit form of the differential interaction rate for a particular detector measured at recoil energy $q$. Let the detector in question be composed of nuclei labeled by the index $j$, each with mass fraction $c_j$. Then, the differential rate of wimp scattering at recoil energy $q$ is\footnote{For a detailed account of \mbox{equation (\ref{rate})} see {\it e.g.} Ref. \cite{Jungman:1995df}.},%\begin{widetext}
\begin{align}
\hspace{-0.25cm}\left.\frac{dR}{dQ}\right|_{Q=q}\!\!\!\!\!=&\frac{2\rho_{\chi}}{\pi m_{\chi}}\!\sum_{j}c_j\!\!\int_{v_{\mathrm{min}_j}(q)}^{\infty}{\!\!\!\!\frac{f(v,t)}{v}dv}\left\{\raisebox{0.5cm}{$\!$}F_{j}^2(q)[Z_jf_p\!+\!(A_j\!-\!Z_j)f_n]^2\!+\!\frac{4\pi}{(2J_j\!+\!1)}\left[a_1^2S_{j_{00}}(q)\!+\!a_0^2S_{j_{11}}(q)\!+\!a_1a_0S_{j_{01}}(q)\right]\!\right\},\label{rate}
\end{align}
%\end{widetext}
where $v_{{\mathrm{min}}_j}(q)$ is the minimum velocity kinematically capable of depositing energy $q$ into the $j^{\mathrm{th}}$ nucleus; $f(v,t)$ is the local velocity distribution function for wimps in the galactic halo; $F_j^2(q)$ and $S_{j_{mn}}(q)$ are nuclear form factor functions for coherent and incoherent scattering, respectively; $Z_j$ and $A_j$ are atomic and mass numbers; $J_j$ is the nuclear spin; $a_1\equiv a_p+a_n$ and $a_0\equiv a_p-a_n$; and the constant parameters $f_{p,n}$ and $a_{p,n}$ describe the coherent and incoherent wimp-nucleon scattering cross sections, respectively\footnote{Imprecisely, coherent scattering is sometimes called `spin-independent' and incoherent scattering `spin-dependent.'}. The interaction parameters can be calculated for any explicit wimp model, regardless of the type of wimp under consideration. \\
\indent In general, this expression for the scattering rate is a second order polynomial in the four (unknown) interaction parameters $f_{p,n}$ and $a_{p,n}$. To highlight this, it can be recast in the suggestive form,%\begin{widetext}
\begin{align}
\hspace{-0.35cm}\left.\frac{dR}{dQ}\right|_{Q=q}\!\!\!\!\!\!=&\frac{2\rho_{\chi}}{\pi m_{\chi}}\!\left\{\!f_p^2\!\left(\sum_{j}\!c_j\int_{v_{\mathrm{min}_j}(q)}^{\infty}{\!\!\!\!\!\frac{f(v,t)}{v}dv}F_j^2(q)Z_j^2\right)\!\!+\!a_p^2\!\left(\!\!4\pi\!\sum_{j}\!c_j\int_{v_{\mathrm{min}_j}(q)}^{\infty}{\!\!\!\!\!\frac{f(v,t)}{v}dv}\frac{\huge[S_{j_{00}}(q)\!+\!S_{j_{11}}(q)\!+\!S_{j_{01}}(q)\huge]}{2J_j+1}\!\right)\right.\nonumber\\
&\!+\!f_n^2\!\left(\!\sum_{j}\!c_j\int_{v_{\mathrm{min}_j}(q)}^{\infty}{\!\!\!\!\!\frac{f(v,t)}{v}dv}F_j^2(q)(A_j\!-\!Z_j)^2\!\right)\!\!+\!a_n^2\!\left(\!\!4\pi\!\sum_{j}\!c_j\int_{v_{\mathrm{min}_j}(q)}^{\infty}{\!\!\!\!\!\frac{f(v,t)}{v}dv}\frac{\huge[S_{j_{00}}(q)\!+\!S_{j_{11}}(q)\!-\!S_{j_{01}}(q)\huge]}{2J_j+1}\!\right)\nonumber\\
&\!\left.+\!f_pf_n\!\left(\!\!2\sum_{j}\!c_j\int_{v_{\mathrm{min}_j}(q)}^{\infty}{\!\!\!\!\!\frac{f(v,t)}{v}dv}F_j^2(q)Z_j(A_j\!-\!Z_j)\!\right)\!\!+\!a_pa_n\!\left(\!\!8\pi\!\sum_{j}\!c_j\int_{v_{\mathrm{min}_j}(q)}^{\infty}{\!\!\!\!\!\frac{f(v,t)}{v}dv}\frac{\huge[S_{j_{00}}(q)\!-\!S_{j_{11}}(q)\huge]}{2J_j+1}\!\right)\!\right\}.\label{rate2}
\end{align}
%\end{widetext}
It is clear from the expression above that using data from\vspace{-0.2cm}
\begin{enumerate}
\item different detector materials (varying the mass fractions, nuclear form factors, nuclear spins, and minimum velocities) and%\vspace{-0.2cm}
\item different recoil energies (varying the nuclear form factors and minimum velocities),
\end{enumerate}
one can invert \mbox{equation (\ref{rate2})} to solve for $\sqrt{\rho_{\chi}}f_{p,n}$ and $\sqrt{\rho_{\chi}}a_{p,n}$ if the halo velocity distribution and $m_{\chi}$ were known. That is, given a halo model and wimp mass, one can use data from different detector materials and different recoil energies to determine $\sqrt{\rho_{\chi}} f_{p,n}$ and $\sqrt{\rho_{\chi}}a_{p,n}$ (up to quadratic ambiguities).\\
\indent Although both the coherent and incoherent (scaled) interaction parameters are in principle calculable, it may not be possible to determine them all when wimps are first discovered. It is perhaps even likely that there will not exist enough data initially to solve for both. For example, if wimps are observed with detectors largely insensitive to spin effects---as is the case for $^{76}$Ge-enriched detectors---then it will not be possible to determine the spin-dependent contribution to scattering.\\
\indent Notice that in \mbox{equation (\ref{rate2})}, the coefficients of $a_{p}$ and $a_{n}$ are linearly independent combinations of the incoherent form factor functions $S_{j_{kl}}(q)$; this is in contrast to $f_{p}$ and $f_{n}$ which are each multiplied by the same form factor function $F_j^2(q)$. This implies that rates observed at different energies can be used to separate $a_p$ from $a_n$ with any single detector capable of measuring incoherent scattering, but rates observed at different energies cannot in general distinguish $f_p$ from $f_n$. It is apparent that to isolate coherent scattering contributions from neutrons and protons one must use data from detectors made of distinct materials.\\
\indent It has not escaped the attention of the authors that knowing the interaction parameters up to an unknown factor---$\sqrt{\rho_{\chi}}$---can give enormous insight into the particle physics of $\chi$. For example, if $\chi$ is the LSP then the ratios $a_p/a_n$ and $a_p/f_n$ could lead to important constraints on $\tan\beta$, the degeneracy of squark masses and mixing parameters, and perhaps contain other information as well. This could be very important for interpreting collider physics and disentangling the MSSM.
\vspace{-0.2cm}
\subsection{Determining the Mass of the Wimp}\begin{figure}[t]\includegraphics[scale=0.7005]{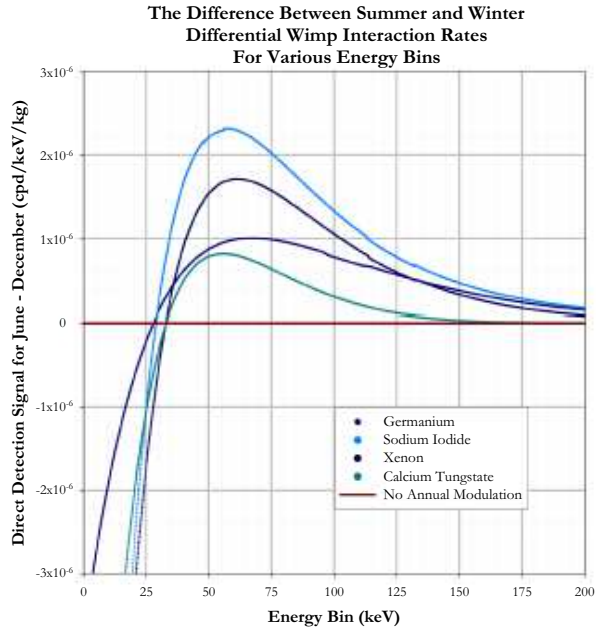}\caption{The difference between direct detection signals in June and December as a function of recoil energy for several detector types. This plot was generated for a model with $m_{\chi}\sim161$ GeV.}\label{annual_modulation}\end{figure}
\indent For a long time there was only one known way to determine the mass of a wimp using direct detection experiments. As the earth orbits the sun, its velocity through the galactic dark matter halo varies between roughly $190$ and $250$ km/s \cite{Munoz:2003gx}. This causes annual modulation in the scattering rate. It has long been known that the amplitude of this annual modulation can change sign between high and low recoil energies; if so, there is a particular energy, the `crossing energy,' at which no annual modulation is observed. This can be seen in \mbox{Figure \ref{annual_modulation}} for several detector materials. In a review article in 1988, Ref. \cite{Primack:1988zm}, Primack {\it et. al.} showed that the crossing energy is directly related to the wimp mass\footnote{Although it seems unlikely to have originated in the review article, we have been unable to find any previous author mentioning the effect. For recent and more thorough descriptions see Refs. \cite{Lewis:2003bv,Green:2003yh}.}. The method is generally robust if the crossing energy is in fact observed.\\ \begin{figure}[t]\includegraphics[scale=0.6877]{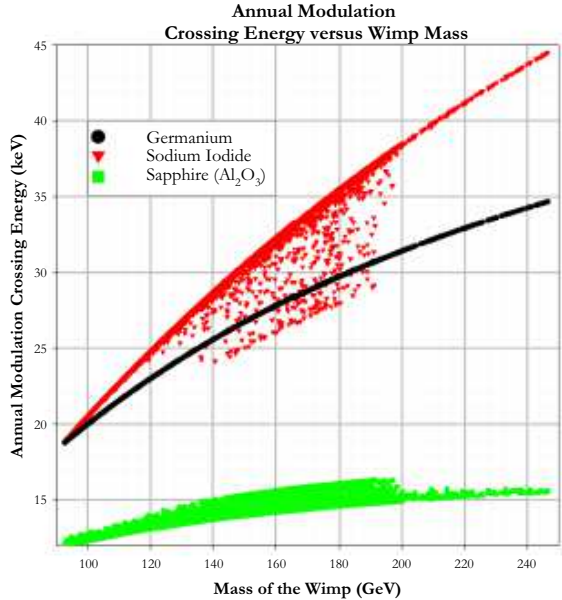}\caption{The annual modulation crossing energy as a function of wimp mass for several thousand randomly-generated constrained MSSMs. Notice that the effect would be difficult to observe for light wimps or with a detector composed of light nuclei, {\it e.g.} sapphire (Al$_2$O$_3$). Notice the ambiguity which arises in the case of detectors which are composed of multiple nuclei.}\label{crossing_all}\end{figure}
\indent Unfortunately, depending on the masses of the wimp and detector nuclei, the crossing energy may not be observable in realistic experiments. Also, the effect can be complicated by the presence of multiple nuclei types in the detector. If the wimp is light then the crossing energy could be lower than any planned detector's threshold: effectively, the annual modulation will peak in June for all of the detector's energy bins (including the lowest)\footnote{Notice that this is precisely what has been observed in the DAMA experiment \cite{Bernabei:2000qi}. Hence, if the DAMA signal is in fact a signal, it favors a (surprisingly) low wimp mass. However, only an upper bound is obtained using this observation alone because the experiment did not observe an energy bin without annual modulation.}. This is also the case for detectors made of relatively light nuclei, {\it e.g.} sapphire (Al$_2$O$_3$). In \mbox{Figure \ref{crossing_all}} we plot the annual modulation crossing energy as a function of mass for detectors made of sodium iodide, germanium, and sapphire. Notice that each of the limitations just described is clearly illustrated in the figure.\\
\indent Recently, we found an alternative technique to determine the mass of a discovered wimp using data from direct detection experiments alone \cite{Bourjaily:2004aj,Bourjaily:2004fp}. This method is based almost entirely on kinematics and will work in principle regardless of what particle the wimp is.\\
\indent The technique presented below may not be entirely unlike those used implicitly in existing analyses (see {\it e.g.} Ref. \cite{Bernabei:2001ve}). However, we are unaware of any formal or explicit presentation of these techniques in the literature.

\subsubsection{Using Kinematical Consistency to Determine the Wimp Mass from Direct Detection Data}
\indent We have found that the wimp mass can be determined if signals are observed in either a single detector that is capable and sensitive enough to observe the incoherent contribution to the scattering rate in several energy bins, or in at least two detectors made of distinct materials in several energy bins. This is done as follows. As we have shown, combining data from multiple detectors and/or from multiple recoil energy bins allows one to isolate the relative scattering contributions from coherent and incoherent interactions with neutrons and protons. To do this, however, the mass of the wimp and the velocity profile were required. If the halo velocity distribution were known, then only the wimp mass would be required. Let us assume this to be the case. Then, given the wimp mass, equation (\ref{rate2}) can be inverted to solve for $\sqrt{\rho_{\chi}}f_{p,n}$ and/or $\sqrt{\rho_{\chi}}a_{p,n}$. Because many direct detection experiments observe scattering rates for many recoil energies, we can expect the system of equations to be over-constrained soon after wimps are first observed.\\
\indent But what if $m_{\chi}$ is not known? The system could still be inverted with an assumed mass $m_{\chi}'$; but if $m_{\chi}'\neq m_{\chi}$ then distinct sets of data would generally give inconsistent results. That is, because the interaction parameters are constants all minimal, linearly independent combinations of measurements used to solve for the scaled interaction parameters will agree if the correct mass were used in the derivation. However, if an arbitrary $m'_{\chi}$ were used to solve for these parameters, different calculations will not in general agree. Therefore, the requirement of consistency in these calculations can be used to identify the correct wimp mass\footnote{This technique is quite similar to $\chi^2$-minimization or maximum likelihood analyses. As stated earlier, similar statistical techniques are already used by experimental groups to determine $m_{\chi}$ (see {\it e.g.} Ref. \cite{Bernabei:2001ve}); but we are not aware of any formal presentation of these in the literature.}. \\ \begin{figure}[t]\includegraphics[scale=0.724]{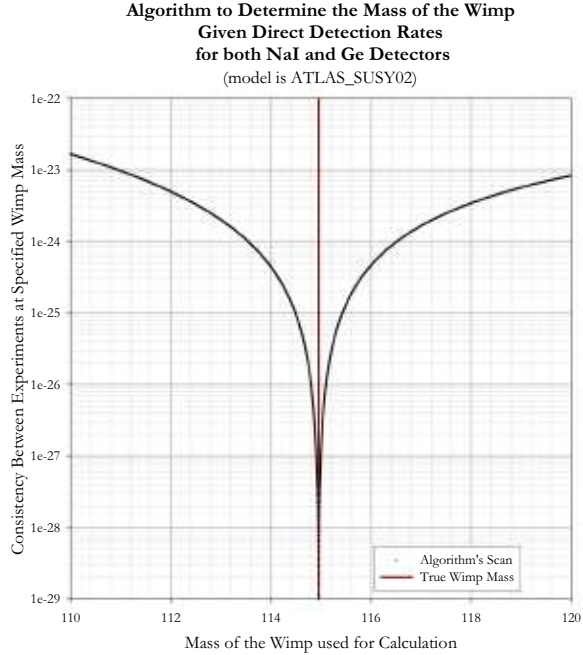}\caption{The function $\zeta(m'_{\chi})$ where the wimp is the neutralino in the MSSM example specified by \mbox{ATLAS} susy point 2 \cite{:1999fr}. The method is equally applicable for arbitrary models of wimp dark matter; in particular, LSP dark matter models with arbitrary soft parameters.}\label{mass_scan}\end{figure} \indent This motivates us to define a `kinematical consistency' function $\zeta(m'_{\chi})$ which compares the values of the scaled interaction parameters obtained using different independent subsets of data as a function of $m'_{\chi}$. Specifically, if there exists sufficient data sufficient data to solve for $\sqrt{\rho_{\chi}}f_{p,n}$\footnote{The analysis is of course improved if both the coherent and incoherent (scaled) parameters are determinable. If so, it should be obvious how to include this information to strengthen the result.}, then let $\zeta(m'_{\chi})$ be given by
\begin{equation}
\zeta(m'_{\chi})\equiv\sum_{i\neq j}\sqrt{\rho_{\chi}}\left\{\left(f_p(i)-f_p(j)\right)^2+\mathrm{similar~terms}\right\},\label{kinematical_consistency}
\end{equation}
where the indices $i,j$ represent distinct minimal sets of data used to compute the constants using the value $m'_{\chi}$ for the wimp mass. It is necessary that $\zeta(m'_{\chi})=0$ when $m'_{\chi}=m_{\chi}$. That is, independent calculations must agree if the correct mass is used. We should mention that this may not be a sufficient condition; but although we have not found any way to demonstrate that $m_{\chi}$ is the unique root of $\zeta(m'_{\chi})$, we have found no example where it has multiple roots. \\
\indent To determine the wimp mass, one varies $m'_{\chi}$ until $\zeta(m'_{\chi})=0$. To test this technique, we applied it to our collection of several thousand randomly-generated, constrained MSSMs. The correct mass was determined by this method for every single model tested. \mbox{Figure \ref{mass_scan}} illustrates a typical plot of $\zeta(m'_{\chi})$, ignoring experimental uncertainties and resolutions. The minimum of $\zeta(m'_{\chi})$ at the correct mass is far lower than $\zeta$ evaluated at ($m_{\chi}\pm5$GeV).\\
\indent There still remain important questions about how this technique can be used in practice. Specifically, our studies have not yet considered the effects of experimental uncertainties, backgrounds, or resolutions. Also, we do not yet have a complete understanding of the requirement that the halo model be known. We have generally found that for direct detection data simulated using a `smooth' halo profile, the correct mass is obtained even if a different (smooth) profile is used in the analysis. However, the effects of using a smooth profile in the analysis if the halo is in fact clumpy or anisotropic is not fully understood.

\section{The Case of Supersymmetric (Neutralino) Dark Matter}
\indent The most popular and perhaps best-motivated wimp candidate for cold dark matter is the lightest supersymmetric particle (LSP) predicted in supersymmetric extensions of the standard model which conserve $R$-parity\footnote{For a comprehensive review of supersymmetric dark matter see Ref. \cite{Jungman:1995df}.}. Indeed, supersymmetric dark matter was predicted well before it was known that non-baryonic dark matter was needed to solve the dark matter problem. In many MSSMs allowed by experimental constraints the LSP is the neutralino, which we denote $\chi$; it is the supersymmetric partner of the neutral gauge and higgs bosons.\\
\indent Based on the arguments presented above, if the LSP is observed in direct detection experiments and produced at colliders then its actual local density can be determined if its scattering cross section ({\it i.e.} its interaction parameters) were known. Similarly, the local density of $\chi$ can be bounded if bounds on the interaction parameters were known. As it turns out, one can bound the parameters using even only sparse data from colliders, without making any assumptions about the MSSM lagrangian parameters. In this section, we will derive this bound explicitly for the most general MSSM.

\newpage
\subsection{Neutralino Interaction Parameters}
\indent Dark matter experiments can be used to determine the effective $\chi$-nucleon interaction parameters $\sqrt{\rho_{\chi}}a_{p,n}$ and/or $\sqrt{\rho_{\chi}}f_{p,n}$. These are of course functions of the $\chi$-quark (and $\chi$-gluon) scattering cross sections. To tree level, the $\chi$-quark scattering cross sections are given in terms of the following diagrams.\\

\begin{center}\begin{fmffile}{issp1}
\begin{fmfchar*}(30,20)
\fmfset{dot_size}{3thin}
\fmfset{arrow_ang}{12}
\fmfleft{i1,i2}
\fmfright{o1,o2}
\fmf{fermion}{i1,v1}
\fmf{fermion}{v1,i2}
\fmf{photon,label=$Z^0$,label.side=right}{v1,v2}
\fmf{fermion}{o1,v2}
\fmf{fermion}{v2,o2}
\fmflabel{$\chi$}{i1}
\fmflabel{$\chi$}{i2}
\fmflabel{$q_i$}{o2}
\fmflabel{$q_i$}{o1}
\end{fmfchar*}
\end{fmffile}\hspace{0.132cm}\raisebox{0.905cm}{$~$}\hspace{0.132cm}\begin{fmffile}{issp2}
\begin{fmfchar*}(30,20)
\fmfset{dot_size}{3thin}
\fmfset{arrow_ang}{12}
\fmfleft{i1,i2}
\fmfright{o1,o2}
\fmf{fermion}{i1,v1}
\fmf{fermion}{v1,i2}
\fmf{dashes,label=$\tilde{q}_j$,label.side=right}{v1,v2}
\fmf{fermion}{o1,v2}
\fmf{fermion}{v2,o2}
\fmflabel{$\chi$}{i1}
\fmflabel{$\chi$}{o2}
\fmflabel{$q_i$}{i2}
\fmflabel{$q_i$}{o1}
\end{fmfchar*}
\end{fmffile}\hspace{0.132cm}\raisebox{0.905cm}{$~$}\hspace{0.132cm}\begin{fmffile}{issp3}
\begin{fmfchar*}(30,20)
\fmfset{dot_size}{3thin}
\fmfset{arrow_ang}{12}
\fmfleft{i1,i2}
\fmfright{o1,o2}
\fmf{fermion}{i1,v1}
\fmf{phantom}{v1,i2}
\fmf{dashes,label=$\tilde{q}_j$,label.side=right}{v1,v2}
\fmf{fermion}{o1,v2}
\fmf{phantom}{v2,o2}
\fmffreeze
\fmf{fermion}{v1,o2}
\fmf{fermion,rubout=3}{v2,i2}
\fmflabel{$\chi$}{i1}
\fmflabel{$\chi$}{i2}
\fmflabel{$q_i$}{o2}
\fmflabel{$q_i$}{o1}
\end{fmfchar*}
\end{fmffile}\hspace{0.132cm}\raisebox{0.905cm}{$~$}\hspace{0.132cm}\begin{fmffile}{issp4}
\begin{fmfchar*}(30,20)
\fmfset{dot_size}{3thin}
\fmfset{arrow_ang}{12}
\fmfleft{i1,i2}
\fmfright{o1,o2}
\fmf{fermion}{i1,v1}
\fmf{fermion}{v1,i2}
\fmf{dashes,label=$h,,H$,label.side=right}{v1,v2}
\fmf{fermion}{o1,v2}
\fmf{fermion}{v2,o2}
\fmflabel{$\chi$}{i1}
\fmflabel{$\chi$}{i2}
\fmflabel{$q_i$}{o2}
\fmflabel{$q_i$}{o1}
\end{fmfchar*}
\end{fmffile}
\end{center}~\\
Given a completely specified MSSM lagrangian it is not difficult to evaluate all of these diagrams and determine each of the interaction parameters. But these will depend on many of the model parameters; in particular, they are functions of
\begin{enumerate}
\item the gauge content of $\chi$ (its photino, zino, and higgsino components),\vspace{-0.2cm}
\item many squark masses and mixing angles,\vspace{-0.2cm}
\item $\tan\beta$, the ratio of the vacuum expectation values of the two higgs doublets,\vspace{-0.2cm}
\item higgs mass parameters (only for the coherent scattering).
\end{enumerate}
It must be emphasized that most of these parameters will be very difficult to measure in practice (especially at hadron colliders). Today, there do not exist general, model-independent methods of determining most of the required parameters at hadron colliders, which is all that will be available for well over a decade.\\
\indent Therefore even years of excellent collider data may be insufficient to properly calculate these scattering cross sections without making unjustified assumptions about supersymmetry breaking, for example. Although it may be tempting to interpret any initial collider data in the limited framework of particular supersymmetry breaking scenario, {\it e.g.} mSUGRA, this can be very misleading. Such analyses trade parameters to purchase an inherent lack of authority. And the possibility of understanding dark matter is too important to oversimplify. \\
\indent However, as we will show, it {\it is} possible to obtain authoritative limits on the LSP-nucleon scattering cross sections without resorting to any model for supersymmetry breaking. More astonishingly, this can be done using only extremely limited data from colliders.

\subsection{An Absolute Bound on the Incoherent Scattering Parameters of the LSP for an Arbitrary MSSM (Using Minimal Input from Colliders)}
\indent Using only a lower limit on the mass of the lightest squark and bounds on $\tan\beta$, we find a strict, model-independent upper bound on the incoherent $\chi$-quark interaction parameters $a_{u,d,s}$. This translates into a strict upper bound on the parameters $a_{p,n}$; together with direct detection data to solve for $\sqrt{\rho_{\chi}}a_{p,n}$, this bound determines a {\it strict, model-independent lower bound on the local density of neutralino dark matter}.\\
\indent Recall that $a_{p,n}$ are functions (and in fact linear combinations) of the $\chi$-quark interaction parameters $a_{u,d,s}$\footnote{The explicit relationships between $a_{p,n}$ and $a_{u,d,s}$ are given in Ref. \cite{Jungman:1995df}.}. Let us derive a strict upper bound for $a_{u}$. Although all three $\chi$-quark parameters are required to compute $a_{p,n}$, it will be obvious how arguments similar to those for $a_{u}$ can be made for $a_{d,s}$.\\
\indent In terms of the lagrangian parameters of an arbitrary softly-broken MSSM, the axial-vector couplings of $\chi$ to the $u,d,s$-quarks are given by\footnote{These expressions are given in terms of the conventions of Ref. \cite{Jungman:1995df}; please refer to its appendix for a more detailed account.}
\begin{align*}
a_u&\!=\!-\frac{g^2}{16m_W^2}(N_{\tilde{H}_1}^2\!-\!N_{\tilde{H}_2}^2)\!+\!\frac{1}{8}\sum_{\tilde{q}_j}\frac{(X_{u1j})^2\!+\!(W_{u1j})^2}{m_{\tilde{q}_j}^2\!-\!(m_{\chi}\!+m_u)^2},\\
a_{d,s}&\!=\!+\frac{g^2}{16m_W^2}(N_{\tilde{H}_1}^2\!-\!N_{\tilde{H}_2}^2)\!+\!\frac{1}{8}\sum_{\tilde{q}_j}\frac{(X_{d1j})^2\!+\!(W_{d1j})^2}{m_{\tilde{q}_j}^2\!-\!(m_{\chi}\!+m_{d,s})^2},\\
\end{align*}
where
\begin{widetext}
\begin{align*}
X_{u1j}&=-g\sqrt{2}\left(\frac{1}{2}N_{\tilde{W}}^*+\frac{1}{6}\tan\theta_WN_{\tilde{B}}^*\right)\left(\Pi_L\Theta_u\right)_{1j}-\frac{gm_{u}}{\sqrt{2}m_W\sin\beta}N_{\tilde{H}_2}^*\left(\Pi_R\Theta_u\right)_{1j},\\
X_{d1j}&=-g\sqrt{2}\left(-\frac{1}{2}N_{\tilde{W}}^*+\frac{1}{6}\tan\theta_WN_{\tilde{B}}^*\right)\left(\Pi_L\Theta_d\right)_{1j}-\frac{gm_{d}}{\sqrt{2}m_W\cos\beta}N_{\tilde{H}_1}^*\left(\Pi_R\Theta_d\right)_{1j},\\
W_{u1j}&=g\sqrt{2}\frac{2}{3}\tan\theta_WN_{\tilde{B}}^*\left(\Pi_R\Theta_u\right)_{1j}-\frac{gm_u}{\sqrt{2}m_W\sin\beta}N_{\tilde{H}_2}^*\left(\Pi_L\Theta_u\right)_{1j},\\
W_{d1j}&=-g\sqrt{2}\frac{1}{3}\tan\theta_WN_{\tilde{B}}^*\left(\Pi_R\Theta_d\right)_{1j}-\frac{gm_d}{\sqrt{2}m_W\sin\beta}N_{\tilde{H}_1}^*\left(\Pi_L\Theta_d\right)_{1j}.
\end{align*}
\end{widetext}
In the expressions above: $\Pi_{L,R}$ are $3\times 6$ projection matrices given in the basis $(\tilde{u}_L,\tilde{c}_L,\tilde{t}_L,\tilde{u}_R,\tilde{c}_R,\tilde{t}_R)$; $\Theta_{u}$ is the unitary matrix which diagonalizes $\tilde{M}_{u}^2$ so that $\tilde{M}^{2~\mathrm{diag}}_{u}=\Theta_{u}^{\dag}\tilde{M}_{u}^2\Theta_{u}$\footnote{We have chosen the basis for the squark-mass matrices so that $M_{u,d}$ are diagonal.}; the subscript $j$ on $\tilde{q}_j$ corresponds to the flavor and handedness of the squarks, so that $j=1,\ldots,6$ corresponds to $(\tilde{u}_L,\tilde{c}_L,\tilde{t}_L,\tilde{u}_R,\tilde{c}_R,\tilde{t}_R)$; and gauge content of $\chi$ is given by $\chi=N_{\tilde{B}}|\tilde{B}\rangle+N_{\tilde{W}}|\tilde{W}\rangle+N_{\tilde{H}_1}|\tilde{H}_1\rangle+N_{\tilde{H}_2}|\tilde{H}_2\rangle$\footnote{As explained briefly before, the fields $\tilde{B}$ and $\tilde{W}$ are the supersymmetric partners of the standard model $U(1)$ and $U(2)$ neutral gauge bosons. They are linear combinations of the supersymmetric partners of the photon and $Z$.}.

\indent Expanding the expression for $a_u$, we obtain%\begin{widetext}
\begin{align}
\hspace{-2cm}a_u=&-\frac{g^2}{16m_W^2}(N_{\tilde{H}_1}^2-N_{\tilde{H}_2}^2)+\frac{g^2}{8}\sum_{\tilde{q}_j}\frac{1}{m_{\tilde{q}_j}^2-(m_{\chi}+m_u)^2}\left\{\raisebox{0.65cm}{~}\right.2\left(\frac{1}{2}N_{\tilde{W}}^*+\frac{1}{6}\tan\theta_WN_{\tilde{B}}^*\right)^2\left(\Pi_L\Theta_u\right)^2_{1j}\nonumber\\
&+ \frac{m_u}{m_W\sin\beta}\mathfrak{Re}\left[\left(N_{\tilde{H}_2}N_{\tilde{W}}^*+\frac{1}{3}\tan\theta_WN_{\tilde{H}_2}N_{\tilde{B}}^*\right)\left(\Pi_R\Theta_u\right)^*_{1j}\left(\Pi_L\Theta_u\right)_{1j}\right]\nonumber\\
&+\frac{m_u^2}{2m_W^2\sin^2\beta}N_{\tilde{H}_2}^2\left(\Pi_R\Theta_u\right)_{1j}^2+\frac{8}{9}\tan^2\theta_WN_{\tilde{B}}^2\left(\Pi_R\Theta_u\right)_{1j}^2\nonumber\\
&-\frac{4m_u\tan\theta_W}{3m_W\sin\beta}\mathfrak{Re}\left[N_{\tilde{H}_2}N_{\tilde{B}}^*\left(\Pi_L\Theta_u\right)_{1j}^*\left(\Pi_R\Theta_u\right)_{1j}\right]+ \frac{m_u^2}{2m_W^2\sin^2\beta}N_{\tilde{H}_2}^2\left(\Pi_L\Theta_u\right)_{1j}^2\left.\raisebox{0.65cm}{~}\right\}.\label{long_a_u}
\end{align}
%\end{widetext}
%\end{widetext}
Similar expressions describe the scattering with $d,s$-quarks.\\
\indent Consider the contribution to $a_u$ from squark exchange---the terms proportional to $1/(m_{\tilde{q}_{j}}^2-(m_{\chi}+m_u)^2)$. Generally, this will be dominated by scattering via the exchange of $\tilde{u},\tilde{d},\tilde{s}-$squarks\footnote{This statement is of course sensitive to the distribution of squark masses and mixing parameters (unknown today).}. Suppose that there exists a lower bound on the mass of the lightest squark, $m_{\tilde{q}_{\ell}}$---which is of course also a lower bound on the mass of each squark. This does not require that the lightest squark has been found; rather, it requires that there exist no squark which has a mass of less than $m_{\tilde{q}_{\ell}}$---so existing, negative collider results would be sufficient. If this is so, then \[\frac{1}{m_{\tilde{q}_j}^2-(m_{\chi}+m_u)^2}\leq\frac{1}{m_{\tilde{q}_{\ell}}^2-(m_{\chi}+m_u)^2}~\forall~j.\]
If the squark masses are nearly degenerate, as is often the case, then the inequality becomes a good approximation. It must be emphasized that our results do not rely on any assumptions about the relative masses of the squarks. However, the limit will be more stringent for models where all the squarks have similar masses\footnote{If the squark masses are widely separated, then the derived upper bound will significantly overestimate the scattering parameters. However, the bound could be significantly improved if, for example, the non-degeneracy of squarks was a fact known experimentally.}. Substituting $1/(m_{\tilde{q}_{\ell}}^2-(m_{\chi}+m_u)^2)$ for $1/(m_{\tilde{q}_{j}}^2-(m_{\chi}+m_u)^2)$ in the summand of \mbox{equation (\ref{long_a_u})}, we can greatly simplify and limit the expression. Notice that the only $j$-dependence remaining in the expression comes from the squark mass mixing parameters, $\left(\Pi_{L,R}\Theta_{u,d}\right)_{1j}$. But here, we recall that unitarity implies \[\sum_{j=1}^6\left(\Pi_{L,R}\Theta_{u,d}\right)_{1j}^2=1\quad\mathrm{and}\quad\sum_{j=1}^6\frak{Re}\left[\left(\Pi_L\Theta_u\right)_{1j}^*\left(\Pi_R\Theta_u\right)_{1j}\right]\leq1.\]
\indent Therefore, we obtain the strict upper bound,%\begin{widetext}
\begin{align*}
\hspace{-0.15cm}a_u\leq&-\frac{g^2}{16m_W^2}(N_{\tilde{H}_1}^2-N_{\tilde{H}_2}^2)+\frac{g^2}{8}\frac{1}{m_{\tilde{q}_{\ell}}^2-(m_{\chi}-m_u)^2}\sum_{\tilde{q}_j}\left\{\raisebox{0.65cm}{~}\right.2\left(\frac{1}{2}N_{\tilde{W}}^*+\frac{1}{6}\tan\theta_WN_{\tilde{B}}^*\right)^2\left(\Pi_L\Theta_u\right)^2_{1j}\\
&+ \frac{m_u}{m_W\sin\beta}\mathfrak{Re}\left[\left(N_{\tilde{H}_2}N_{\tilde{W}}^*+\frac{1}{3}\tan\theta_WN_{\tilde{H}_2}N_{\tilde{B}}^*\right)\left(\Pi_R\Theta_u\right)^*_{1j}\left(\Pi_L\Theta_u\right)_{1j}\right]\\
&+\frac{m_u^2}{2m_W^2\sin^2\beta}N_{\tilde{H}_2}^2\left(\Pi_R\Theta_u\right)_{1j}^2+\frac{8}{9}\tan^2\theta_WN_{\tilde{B}}^2\left(\Pi_R\Theta_u\right)_{1j}^2\\
&-\frac{4m_u\tan\theta_W}{3m_W\sin\beta}\mathfrak{Re}\left[N_{\tilde{H}_2}N_{\tilde{B}}^*\left(\Pi_L\Theta_u\right)_{1j}^*\left(\Pi_R\Theta_u\right)_{1j}\right]+ \frac{m_u^2}{2m_W^2\sin^2\beta}N_{\tilde{H}_2}^2\left(\Pi_L\Theta_u\right)_{1j}^2\left.\raisebox{0.65cm}{~}\right\},\\
\leq&-\frac{g^2}{16m_W^2}(N_{\tilde{H}_1}^2-N_{\tilde{H}_2}^2)+\frac{g^2}{8}\frac{1}{m_{\tilde{q}_{\ell}}^2-(m_{\chi}-m_u)^2}\left\{\raisebox{0.65cm}{~}\right.2\left(\frac{1}{2}N_{\tilde{W}}^*+\frac{1}{6}\tan\theta_WN_{\tilde{B}}^*\right)^2\\
&+ \frac{m_u}{m_W\sin\beta}\mathfrak{Re}\left(N_{\tilde{H}_2}N_{\tilde{W}}^*+\frac{1}{3}\tan\theta_WN_{\tilde{H}_2}N_{\tilde{B}}^*\right)+\frac{m_u^2}{2m_W^2\sin^2\beta}N_{\tilde{H}_2}^2+\frac{8}{9}\tan^2\theta_WN_{\tilde{B}}^2\\
&-\frac{4m_u\tan\theta_W}{3m_W\sin\beta}\mathfrak{Re}\left(N_{\tilde{H}_2}N_{\tilde{B}}^*\right)+ \frac{m_u^2}{2m_W^2\sin^2\beta}N_{\tilde{H}_2}^2\left.\raisebox{0.65cm}{~}\right\},\\
=&-\frac{g^2}{16m_W^2}(N_{\tilde{H}_1}^2-N_{\tilde{H}_2}^2)+\frac{g^2}{8}\frac{1}{(m_{\tilde{q}_{\ell}}^2-(m_{\chi}+m_u)^2}\left\{\raisebox{0.65cm}{$\!$}\frac{17}{18}\tan^2\theta_WN_{\tilde{B}}^2+\frac{1}{2}N_{\tilde{W}}^2+\frac{m_u^2}{m_W^2\sin^2\beta}N_{\tilde{H}_2}^2\right.\\
&+\!\frac{1}{3}\tan\theta_W|N_{\tilde{B}}||N_{\tilde{W}}|\cos(\alpha_{\tilde{W}}\!)\!+\!\frac{m_u}{m_W\sin\beta}|N_{\tilde{W}}||N_{\tilde{H}_2}|\cos(\alpha_{\tilde{H}_2}\!\!-\!\alpha_{\tilde{W}})\!-\!\frac{m_u}{m_W\sin\beta}\tan\theta_W|N_{\tilde{B}}||N_{\tilde{H}_2}|\cos(\alpha_{\tilde{H}_2}\!)\left.\raisebox{0.65cm}{$\!$}\!\!\right\},
\end{align*}
%\end{widetext}
where $\alpha_{\tilde{H}_2}$ $(\alpha_{\tilde{W}})$ is phase between $N_{\tilde{B}}$ and $N_{\tilde{H}_2}$ $(N_{\tilde{W}})$\footnote{Note that we have allowed the soft lagrangian parameters to be complex.}. \\
\indent This expression has seven {\it real} unknowns including $\sin\beta$. By the normalization of the neutralino wave function, $|N_{\tilde{B}}|^2+|N_{\tilde{W}}|^2+|N_{\tilde{H}_1}|^2+|N_{\tilde{H}_2}|^2=1,$ the six unknowns \mbox{$\{|N_{\tilde{B}}|,|N_{\tilde{W}}|,|N_{\tilde{H}_1}|,|N_{\tilde{H}_2}|,\alpha_{\tilde{W}},\alpha_{\tilde{H}_2}\}$} form a compact parameter space. Therefore, $a_{u}$ can be absolutely maximized with respect to all these numbers (unknown today). The only unknown which cannot be marginalized is $\sin\beta$. To maximize $a_u$ with respect to $\sin\beta$, a lower bound of $\sin\beta$ must exist---corresponding to a lower bound of $\tan\beta$\footnote{Both the upper and lower bounds of $\tan\beta$ are important. This is because a lower bound of $\cos\beta$---effectively an upper bound on $\tan\beta$---is needed for $a_{d,s}$.}. Using this bound in place of $\sin\beta$ in the expression above, we obtain an absolute upper bound for $a_u$.\\
\indent Notice that in our work above we allowed all of the gauge content of the neutralino to be unknown and admitted possibly complex, arbitrary soft parameters and phases. In short, {\it the analysis applies to the most general softly-broken supersymmetric standard model}.\\
\indent It is important to note the flexibility and improvability of the derivations for these bounds. If, for example, the masses of several light squarks were known then one could greatly improve the bounds by including these in the explicit expression for $a_{q}$ and maximizing it relative to the remaining unknown parameters. In this manner almost any additional knowledge can be added to arrive at stronger results. Therefore, not only do these bounds grow more restrictive with increasing knowledge of the MSSM, but they continue to approach a realistic estimate of the interaction parameters. It should also be noted that the approach is easily generalizable to the case of a (neutralino) metastable next-to-lightest supersymmetric particle (NLSP) if it is the dark matter observed in these experiments.\\
\indent The coherent parameters are similar to the incoherent ones except that they also contain higgs exchange at tree-level. This implies that in addition to squark masses, mixing angles, $\tan\beta$, and gauge content of $\chi$, one must know the higgs masses. Therefore, in general, less is required to compute $a_{p,n}$ than $f_{p,n}$. But even with allowing the higgs masses to be `known,' we have not been able to find robust bounds for the parameters $f_{p,n}$ comparable to those for $a_{p,n}$. It appears that much more data from colliders will be required to compute or bound $f_{p,n}$.

\subsection{A Strong Lower Bound on the Local Density of Neutralino Dark Matter}\vspace{-0.2cm}
\begin{figure}[t]\includegraphics[scale=0.648]{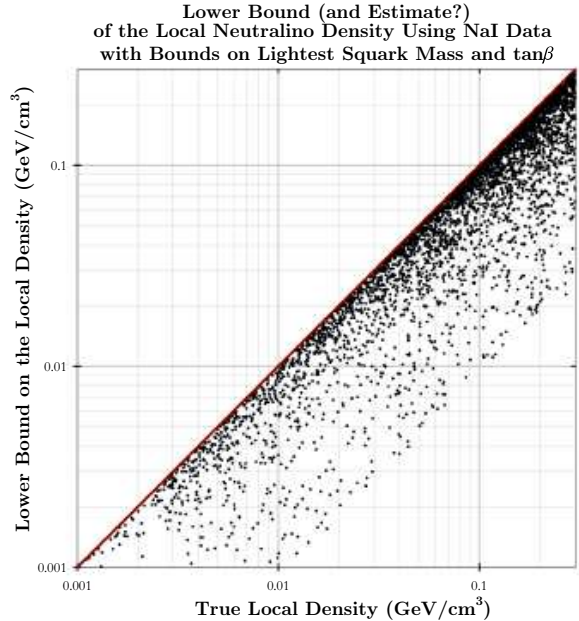}\caption{This plot compares the lower bound (and estimate?) of the neutralino local density computed using the strong upper bound for $a_{p,n}$ to the true density for thousands of models. Notice that the procedure correctly determined a lower bound for the local density for every model. Unfortunately, sometimes the lower bound was well below the true density; this is the case when the lower bound on the lightest squark mass was poor. Recall that the local density of halo dark matter is $\rho_{\mathrm{cdm}}\sim0.3$ GeV/cm$^3$.}\label{density_bound}\end{figure}
\indent In the previous section we showed that even weak bounds on the mass of the lightest squark and the range of $\tan\beta$ allows one to obtain strict, model-independent upper bounds on the incoherent scattering parameters of $\chi$ for neutrons and protons. Because direct detection data measures the product of the scattering parameters and the local density of neutralinos in the halo, the upper bound on $a_{p,n}$ immediately translates into a strong, model-independent lower bound on the local neutralino density.\\
\indent To test this, we considered several thousand randomly-generated, experimentally constrained MSSMs. For each of these models, we calculated the interaction rates for a NaI detector in twelve recoil energy bins. This (idealized) data was used to compute the mass of the LSP, using the kinematical consistency function, and to solve for $\sqrt{\rho_{\chi}}a_{p,n}$. Upper bounds were calculated for $a_{p,n}$ assuming $10\%$ uncertainty in $\tan\beta$ and a lower bound on the lowest squark mass of either $200$ GeV or the actual mass of the lightest squark, whichever was less. The specific gauge content of the neutralino was taken to be known for each model for the sake of computational simplicity\footnote{If the gauge content of the neutralino was unknown, the interaction parameters could have been maximized with respect to these parameters as described earlier. Therefore, the upper bounds obtained are more restrictive than would be in practice.}. Using the upper bounds for $a_{p,n}$, we obtained a lower bound on the local density of neutralino dark matter, $\rho_{\chi}$.\\
\mbox{Figure \ref{density_bound}} illustrates the results of using this algorithm for each of the randomly generated MSSMs. Notice that the estimated local density is always strictly less than the true local density---as required by it being a lower bound. Also, notice that for many models the lower bound was not such a poor estimate. This will be the case, for example, when the lightest squark mass is near or below the (assumed to be $200$ GeV) lower bound and the squarks are relatively degenerate.

\section{Conclusions and Discussion}\vspace{-0.2cm}
\indent We have seen from general arguments and robust specific examples that a discovery of dark matter particles in the galactic halo or at colliders cannot by itself solve the dark matter problem. To determine the cosmological significance of discovered wimps, their local or relic density must be determined. We have proved by example that this cannot be done using data from colliders alone or in/direct detection experiments alone. However, we have shown that it may be possible to determine the actual local density of discovered wimps by combining data from both colliders and detectors: detectors to measure the product of the local density and the scattering cross section, and colliders to estimate or bound the scattering cross section. Up to specific and possibly resolvable astrophysical uncertainties regarding the homogeneity and isotropy of the local dark matter halo, this may be sufficient to solve the local dark matter problem.\\
\indent To determine their local density, not only must wimps be observed both in direct detection experiments and at colliders, but it must be established that these two signals correspond to the same particle. The only known way to suggest that the same particle is observed in both detectors and colliders is to compare the masses of the particles seen in each. This requires that the wimp mass be determined independently by both direct detection experiments and colliders. We reviewed the heretofore known way to determine the wimp mass using direct detection experiments, and presented a somewhat new, explicit technique. We discussed how, in contrast to direct detection experiments, there does not exist any general method of determining the mass of wimps produced at hadron colliders, though it may be possible in particular lucky cases.\\
\indent To highlight the general arguments presented, we often referred to the case of supersymmetric dark matter, where the discovered wimp is the LSP. We criticized the trend of studying dark matter by calculating the thermal freeze-out contribution to the LSP relic density. We showed that determining even the (quintessence-free) thermally-produced contribution to the relic density of the LSP is both fundamentally insufficient and likely a practical impossibility using data from hadron colliders. In general, too many (possibly) difficult to measure lagrangian parameters are required to compute this for a generic supersymmetric model; and even if all the parameters were known, the result can be extremely sensitive to high-scale physics unobservable at colliders.\\
\indent In contrast to thermal freeze-out calculations, we found that the LSP-nucleon incoherent scattering cross section could be absolutely bounded (and perhaps well-approximated) using only sparse data from colliders for the most general MSSM (without any presumptions about supersymmetry breaking); furthermore, we showed that this bound is iteratively improvable as more data becomes available from colliders. This bound was shown to be robust for thousands of realistic test models. Combining this bound on the scattering cross section with data from direct detection experiments immediately allows one to obtain into an absolute lower bound on the local density of neutralino dark matter.\\
\indent This is the first study of which we are aware that fully addresses the difficulty of determining what fraction of dark matter is accounted for by wimps discovered at colliders or in detectors. We have shown explicitly how to deal with this problem in the case of supersymmetric dark matter, and described how similar analyses could be done for any other dark matter particle candidate. Although the dark matter problem may not be solved immediately when wimps are first observed, there are clear and general ways to address their cosmological significance. \\\vspace{-0.2cm}

\noindent{\bf Acknowledgements}\\
\indent We are grateful to Frank Paige for many helpful discussions about supersymmetry phenomenology at the LHC, and to Brent Nelson, Anne Green, Liantao Wang, Ting Wang, and Chris Savage for helpful criticism and comments on drafts of this work. This research was supported by US Department of Energy, the Michigan Center for Theoretical Physics and the National Science Foundation.\\

\end{document}